# High Performance Thin-film Lithium Niobate Modulator Applied ITO Composite Electrode with Modulation Efficiency of 1V·cm

Xiangyu Meng,[1,&] Can Yuan,[1,&] Xingran Cheng,[1] Shuai Yuan,[1] Chenglin Shang,[1] An Pan,[1] Zhicheng Qu,[1] Xuanhao Wang,[1] Peijie Zhang,[2] Chengcheng Gui,[2] Chao Chen,[1,a)] Cheng Zeng[1,a)] and Jinsong Xia[1,a)]

[1] Wuhan National Laboratory for Optoelectronics, School of Optical and Electronic Information, Huazhong University of Science and Technology, Wuhan 430074, China

[2] Huawei Technologies, B & P Laboratory, Shenzhen 518000, China

[&] These authors contributed equally to this work and should be considered co-first authors.

[a)] Authors to whom correspondence should be addressed:

cchen@hust.edu.cn , zengchengwuli@hust.edu.cn and jsxia@hust.edu.cn

**Abstract:** Thin film lithium niobate (TFLN) based electro-optic modulator is widely applied in the field of broadband optical communications due to its advantages such as large bandwidth, high extinction ratio, and low optical loss, bringing new possibilities for the next generation of high-performance electro-optic modulators. However, the modulation efficiency of TFLN modulators is still relatively low when compared with Silicon and Indium-Phosphide (InP) based competitors. Due to the restriction of the trade-off between half-wave voltage and modulation length, it is difficult to simultaneously obtain low driving voltage and large modulating bandwidth. Here, we break this limitation by introducing Transparent Conductive Oxide (TCO) film, resulting in an ultra-high modulation efficiency of 1.02 V·cm in O-Band. The fabricated composite electrode not only achieves high modulation efficiency but also maintains a high electro-optic bandwidth, as demonstrated by the 3 dB roll-off at 108 GHz and the transmission of PAM-4 signals at 224 Gbit/s. Our device presents new solutions for the next generation of low-cost high-performance electro-optic modulators. Additionally, it paves the way for downsizing TFLN-based multi-channel optical transmitter chips.

# 1. Introduction

High-speed, low half-wave-voltage electro-optic (EO) modulators play a vital role in fiber-optic communication networks[1] and microwave photonics[2], where they are used to transport microwave electrical signals into light signals. The preferred choice for most commercial EO modulators is lithium niobate (LN), due to its high electro-optic coefficient (≈31 pm/V) and transparency at telecommunication wavelengths[1]. However, the fabrication of LN optical waveguides is challenging, resulting in traditional bulk LN modulators relying on titanium-in-diffusion or proton-exchange methods[3, 4] which have a low refractive index contrast of around 0.1. Consequently, these methods lead to weak optical confinement, resulting in large optical mode sizes and gaps between electrodes, ultimately yielding low electro-optic (EO) efficiency.

Lithium niobate on insulator (LNOI), also known as thin-film lithium niobate (TFLN), is considered as a promising platform for the development of photonic integrated circuits (PICs). By dry etching the LN layer, the refractive index contrast of LN ridge waveguides is significantly higher (approximately 0.7) compared to waveguides on bulk LN[5], resulting in small bend radius, compact size and large integration ability. As a result, plenty of photonics devices have been reported on LNOI platform[6-14]. Notably, EO modulators with low half-wave-voltage and large bandwidth have been successfully demonstrated[15-19].

For further applications in commercial scenarios, the improving modulation efficiency of LN modulators is of vital importance. Higher modulation efficiency is desirable as it leads to a smaller driving voltage and lower power consumption. The modulation electrode spacing can be greatly narrowed compared to bulk materials, due to the strong confinement of the optical field by LNOI waveguides with high refractive index contrast. The EO modulation efficiency of TFLN typically ranges from 1.7 to 2.2 V·cm in C-Band[17-26], while it is generally lower in O-Band. Prof. Chu. from Zhejiang University has achieved a high modulation efficiency of 1.41 V·cm in C-Band by utilizing the high dielectric constant cladding in the TFLN platform. However, this efficiency is still lower compared to the SOI (1.35 V·cm)[27] and InP (0.6 V·cm)[28] platforms. This low efficiency results in the longer modulation arm in the TFLN modulator and a larger coverage area than the silicon optical chip, creating obstacles for commercial applications. Consequently, this limitation needs to be addressed.

Recently, single-wave 200 Gbps transmission using high-performance TFLN electro-optic modulators has been demonstrated successfully. However, when compared to differentially driven silicon modulators, LNOI modulators with Ground-Signal-Ground (GSG) electrodes do not offer significant advantages in terms of size, cost, and power consumption. To address this issue, it is important to significantly improve the modulation efficiency of the LNOI electro-optical modulator. This will enable the development of an array modulator with large bandwidth, low half-wave voltage, and small size. By doing so, the cost and power advantages of LNOI optical transmitter chips can be further enhanced, thereby facilitating their future applications in next-generation optical modules.

In this paper, we proposed an ultra-high efficiency EO Mach-Zehnder modulator (MZM) on LNOI platform with the combined electrodes of TCO (Transparent Conductive Oxide) material and gold. The absorption coefficient $k$ of the TCO films is optimized to a low level of 0.001, resulting in a narrow composite electrodes gap of 3 μm with the measured instertion loss of 2.9 dB. This optimization leads to a low half-wave-voltage length product of only 1.024 V·cm, which is the highest record value for MZM type LNOI modulators. We measured a 3-dB EO S21 roll-off at 108 GHz with a 5-mm-long modulation region. Furthermore, we successfully achieved implementation of 112Gbit/s On-Off Keying (OOK) signal transmission and 224 Gbit/s signal transmission using four-level pulse amplitude (PAM-4) modulation.

## 2. Principles and Structure

### 2.1 Design principles

In the X-cut LN based modulator, the half-wave-voltage length product $V_{\pi^*}L$ can be written as:

$$V_\pi * L = \frac{\lambda\, g}{n^3 \gamma_{33} \Gamma} \tag{1}$$

Where $L$ is the length of modulation region, $\lambda$ is the wavelength of the input light, $n$ is the optical refractive index, $\gamma_{33}$ is the electro-optic coefficient of LN, $\Gamma$ is the overlap integral between optical modes and electrical field, and $g$ is the gap of electrodes. According to Equation (1), reducing the gap $g$ is an effective method for achieving high EO efficiency or a low $V_{\pi^*}L$. However, the narrow electrode gap would introduce large optical absorption loss. Fabricating electrodes on an optically transparent buffer layer, such as silica, is a regular method on traditional bulk LN platform to reduce optical loss. Prof. Xiong. from Tsinghua University conducted an experiment on the TFLN platform, where they inserted a 100-nm thick silica buffer layer between the LN waveguide and electrodes. They successfully demonstrated a modulator with a half-wave-voltage length product of only 1.7 in C-band, utilizing 3 μm electrode gap[16].

In order to further reduce the size of the modulator and improve its bandwidth, a new type of low loss electrode is urgently needed. Transparent conducting oxide (TCO) materials have both superior performance of optical transparent and adjustable conductivity and have been applied to EO modulators of other material platforms[29-33]. By adjusting the carrier concentration and mobility of the TCO material, we can construct a composite electrode by combining it with the gold electrode. This composite electrode can be placed closer to the waveguide without causing enhanced optical loss, which is advantageous for higher electro-optic (EO) efficiency. By adjusting the cladding thickness and buffer layer thickness, we can achieve a microwave refractive index that is close to the light group index. This can be achieved through specific structural design. Additionally, maintaining the advantage of high bandwidth is possible by incorporating traveling wave electrodes into the system.

**2.2 Design of TFLN modulator with ITO composite electrode**

The schematic structure of the demonstrated modulator is shown in Fig. 1(a) and 1(b). Two 3-dB multimode interference (MMI) couplers are used to split the input light and combine them after modulation. Ground-Signal-Ground (GSG) electrode structure is used for Push-Pull modulation manner to reduce $V_\pi$. The conductivity of TCO materials ($10^2 \sim 10^5$ S/m) is much less than that of gold (~$10^8$ S/m). According to our simulation, replacing gold completely with TCO material is not useful because of the degradation of microwave performance at high frequency. So, we combine TCO and gold as the travelling microwave electrode to narrow the electrodes gap without degrading the quality of radio frequency (RF) signal transmission. Partially enlarged scanning electron microscope (SEM) image of the combined electrodes is shown in Fig. 1(c).

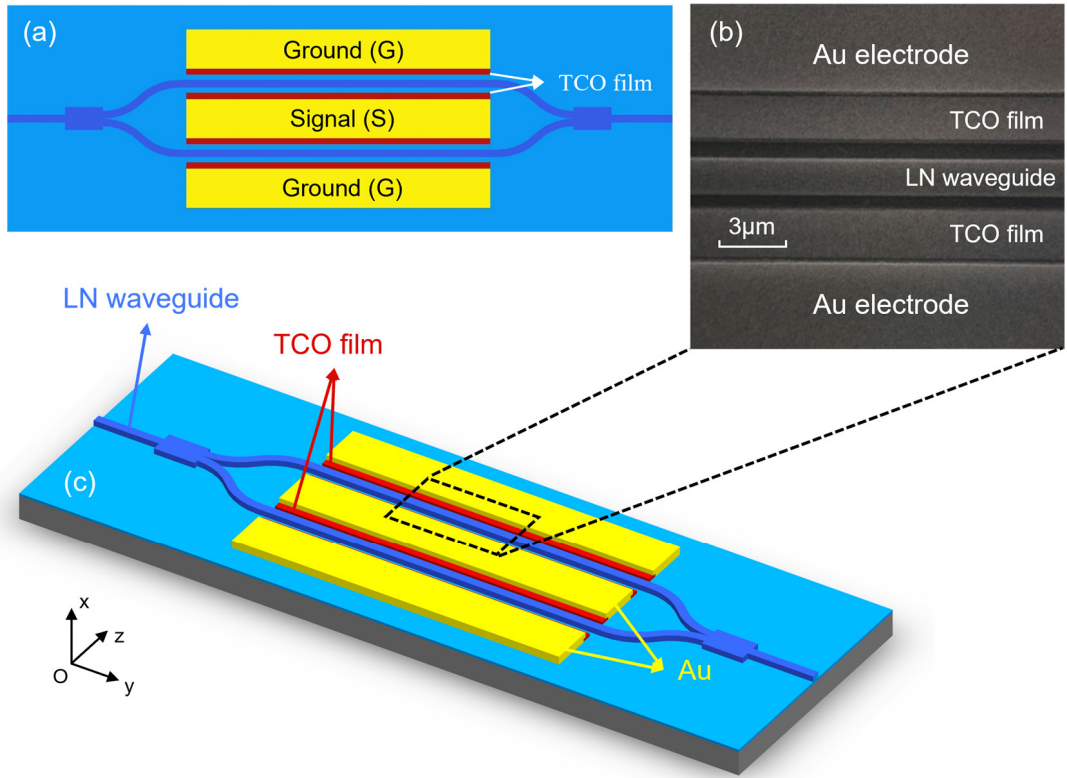

Fig. 1. (a) Top view of the modulator. (b) SEM image of combined electrodes and waveguides. (c) 3D schematic of the modulator.

Indium Tin oxide (ITO) is one of many proposed TCO materials and is suitable for application on LNOI platform. Thanks to its much lower refractive index than LN (1.5 is measured at 1310 nm), ITO doesn't affect the optical modes in waveguides. The imaginary part of refractive index at 1310nm of 0.001 and conductivity of 400 S/m are presupposed respectively in the simulation. Fig. 2 provides a depiction of the optical mode and electrical mode in one arm of the MZM (a phase modulator). In the modulator part of the ridge waveguide, the ridge height ($h$) is 260 nm, the width ($w$) is 2 μm, and the thickness ($s$) of the slab part is 240 nm. The TCO loaded composite electrode is positioned on a 100 nm thick silicon oxide buffer. The gold electrode has a spacing

($d_{Au}$) of 5 μm, while the TCO electrode has a spacing ($d_{TCO}$) of 3 μm. The thicknesses of the gold electrode and TCO electrode, represented as $t_{Au}$ and $t_{TCO}$ respectively, are 1.4 μm and 180 nm. Fig. 2(b) clearly illustrates that the optical mode is tightly confined in waveguides with a narrow gap between the ITO electrodes. Consequently, the electric field applied on the optical modes is significantly increased. Figs. 2(c) and 2(d) display the electric field distribution of the conventional modulator without TCO assistance and the modulator we designed with TCO assistance, respectively. To further demonstrate the significant enhancement in electric field intensity brought about by our designed modulator, we selected a sample point at the center of the waveguide. The simulation results show that the electric field intensity at the center sample point of the two waveguides increased from $1.88 \times 10^5$ V/m to $3.07 \times 10^5$ V/m by incorporating TCO materials to assist in electric field conduction. The contribution of TCO materials in facilitating electric field conduction is substantial.

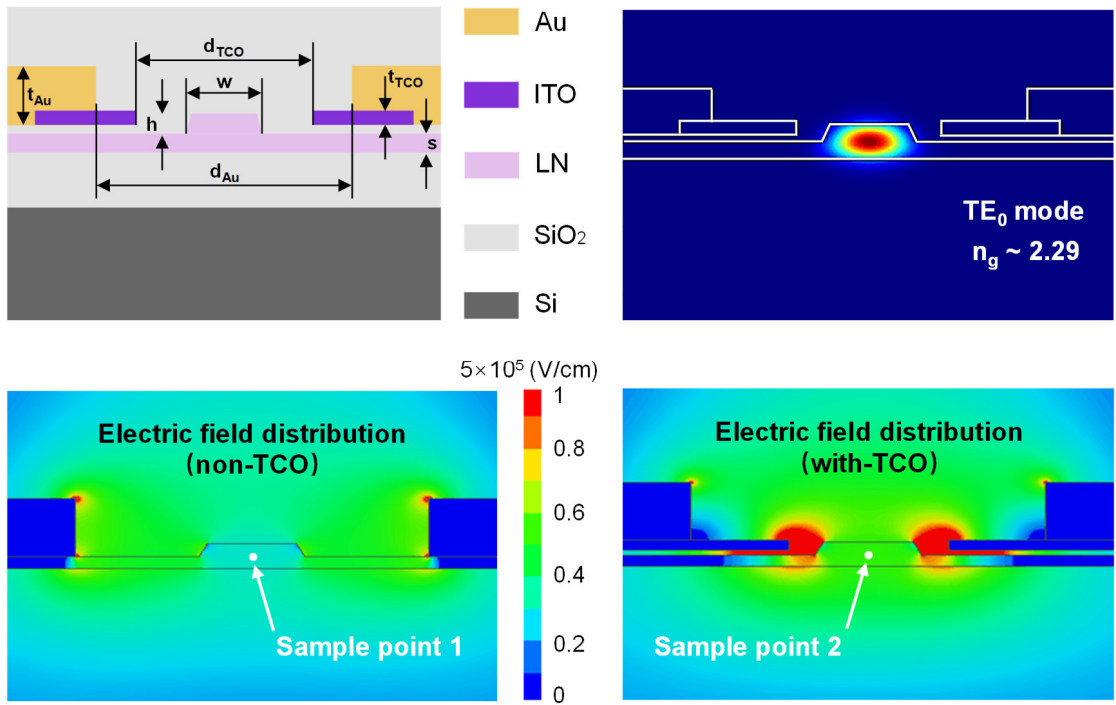

Fig. 2. (a) Cross section of the phase modulator. (b) Optical mode in LNOI waveguides with combined electrodes. With the same drive voltage (1 V), the comparison of the electric field in (c) regular LNOI modulator with 5 μm electrodes gap and in (d) LNOI waveguide with narrow-gap combined electrodes (the gold electrodes gap is 5 μm and the ITO electrodes gap is 3 μm).

**2.3 Structural, electrical and optical properties of the ITO films**

In simulation, the ITO film electrode placed near the waveguide is required to have high optical properties, its extinction coefficient is determined to be in the order of 0.001, and its conductivity should be as high as possible to maintain high microwave performance at high frequency. According to Lorentz-Drude dispersion theory, high concentration of carrier will cause photon dispersion in infrared band, which is the main cause of ITO waveguide loss[34, 35]. Therefore, we need to develop ITO thin films with low carrier concentration and high conductivity. In the deposition process, the actual material properties of ITO are very sensitive to the preparation conditions, and the various defect concentrations, crystallinity and roughness of ITO are determined, which directly determine the final photoelectric properties of ITO films.[36-38] In previous studies, oxygen concentration in the preparation process has been proved to be an important factor affecting the photoelectric performance of ITO film[39]. At high oxygen concentration, oxygen reacts fully with ITO, reducing the oxygen vacancy in the film and decreasing the carrier concentration, and the scattering absorption of ITO film in infrared band is reduced finally. Therefore, in the process of preparing ITO film, the oxygen flow rate has been fixed at highest proportion to obtain a lower loss of ITO electrode film.

In the process of ITO deposition, in addition to the concentration of oxygen, the evaporation and deposition rate of ITO are also important factors affecting the reaction process, which determine the electrical parameters of the film. After optimizing the deposition rate and other process parameters, we successfully prepared high-performance ITO films on quartz substrate and modulator devices at room temperature. As shown in Fig. 3a, the surface information of ITO films is obtained by Atomic Force Microscope (AFM) analysis, and the root mean square average ($R_q$) of height deviation taken from the image is calculated to be 0.4 nm, indicating that the surface of ITO films is very smooth and suitable for further deposition. The optical properties of ITO films are the most important parameters, the transmittance and absorptivity curve of ITO films are obtained by UV-Vis-NIR spectrophotometer and ellipsometer tests (Figs. 3(a) and 3(b)). It could be found the infrared transmittance of ITO films remained above 90%, and the extinction coefficients of ITO films were $k$=0.0011 at 1310 nm and $k$=0.0013 at 1550 nm, respectively, fully meeting the requirements of modulator simulation. Besides, the conductivity of ITO is still maintained a high level of 477 S/m due to the high mobility of 29 cm$^2$/(V·s).

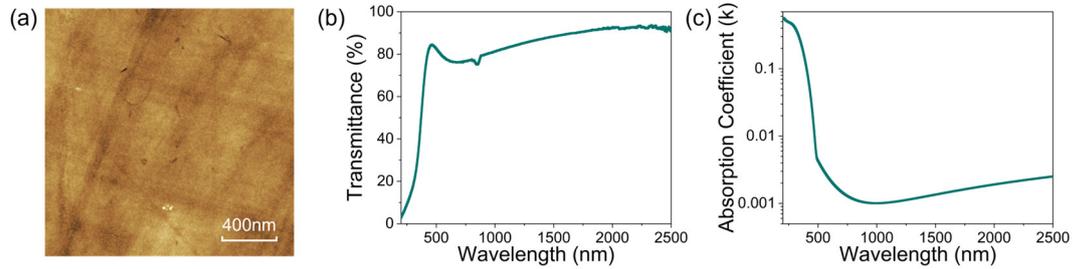

Figure 3. (a) AFM image of the ITO film. (b) The transmittance spectra of ITO film with thickness of 100 nm. (c) The absorption coefficient spectra of ITO film.

## 3. High-speed modulation performance

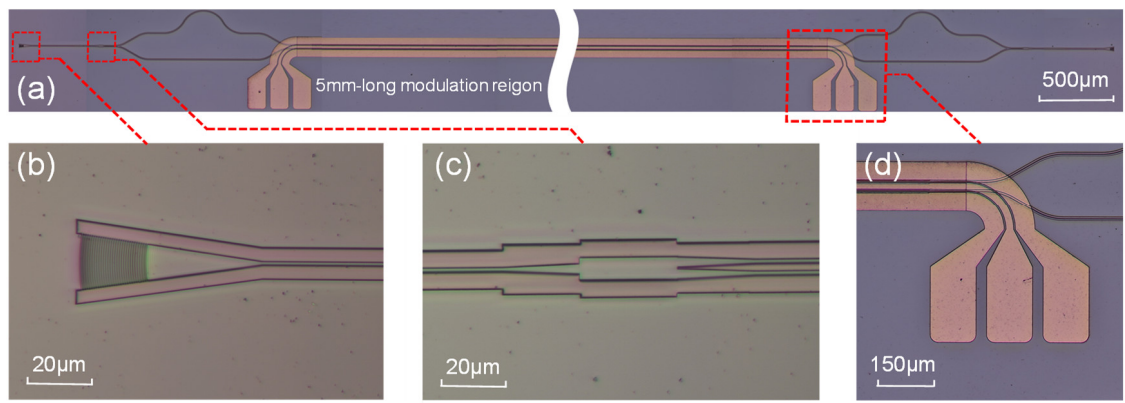

Figure 4. (a) Overview of modulator under microscope; (b) Grating coupler; (c) Input MMI; (d) Terminal electrode pads.

The overall view of the modulator is shown in Fig. 4(a). There are two designed gratings at both ends of the device to couple the light into and out of the chip for measurement in Fig. 4(b). The TE$_0$ mode light is coupled into the input grating, which is equally divided into two beams through the multi-mode interference coupler (MMI) as shown in Fig. 4(c). The modulator is then coupled out at the same grating as the input. The TCO films can be seen under SEM, as shown in Fig. 1(b). Fig. 4(d) shows the terminal electrode pads. During the test, a 23 Ω terminal resistor is connected to the electrode pads through gold wire. The measured insertion loss of the demonstrated modulators with 3-mm TCO gap is 2.9 dB at 1310 nm.

### 3.1 Modulation efficiency performance

The test schematic for the half-wave voltage measurement is shown in Fig. 5(a), where a 100 kHz sawtooth electric signal is applied onto the electrodes. All of the measured devices have a 5-mm-long modulation region. From Fig. 5(b), it can be calculated that the 5-mm-long modulator with 3-μm TCO gap has a half-wave voltage of 2.04 V. Thus, the half-wave-voltage length product $V_\pi*L$ is as low as 1.02 V·cm. It is worth noting that the extra insertion loss caused by the absorption of the combined electrodes of TCO and gold remains at a low level. The measured $V_\pi*L$ of modulators is in great agreement with simulation results as depicted in Fig. 5(c).

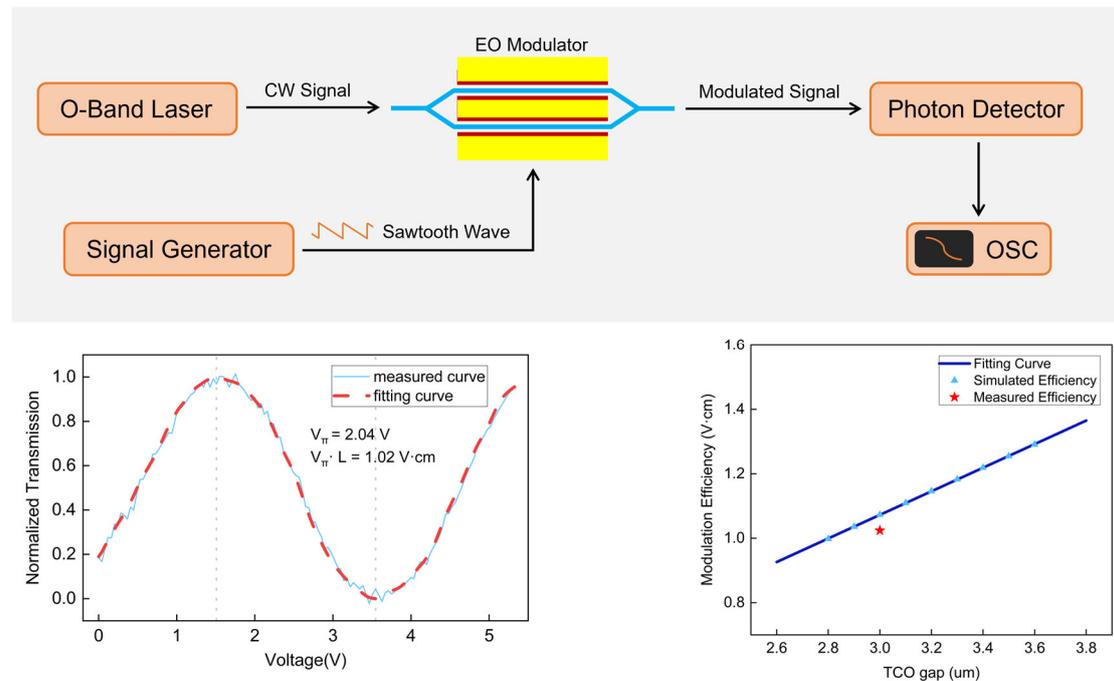

Fig. 5. (a) Normalized optical transmission of the 5-mm modulator with 3-μm TCO gap as a function of the applied voltage. (b) Variation of $V_\pi*L$ with the TCO gap. (c) Modulation efficiency simulation under different TCO-gap with the measured data.

### 3.2 Bandwidth, Eye Diagram and BER performance

In the measurements of bandwidth, a GSG RF probe is used to apply microwave signal onto electrodes. The microwave transmission S21 and the reflections S11 are measured by vector network analyzer (Agilent N5227A, VNA). Using CST Stduio Suite solver (SIMULIA), S21 can be simulated as shown in Fig. 6. We tested the 5-mm-long modulation arm device we created and found that its EO S21 response rolled down to -3dB at 108 GHz, demonstrating impressive bandwidth performance. Additionally, our simulation results closely matched the test results.

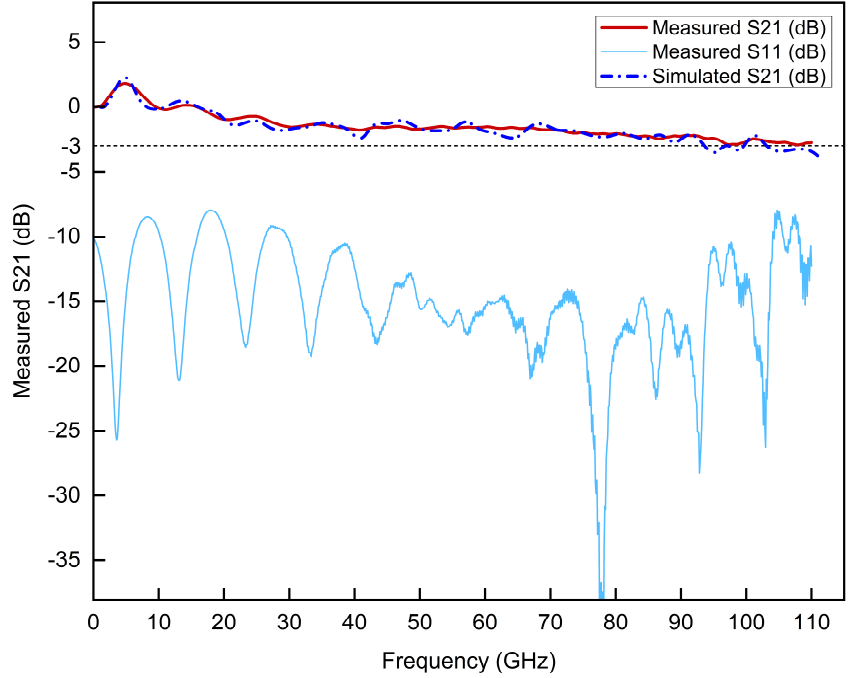

Fig. 6. Electro-optic performance of the 5 mm long LN modulators.

During the test, a 70 GHz arbitrary waveform generator (Keysight M8199A) was used to generate the data, which was then amplified by a Driver (SHF S804B) with an amplification factor of 22dB and connected to the modulator. The output signal was measured by a 110 GHz sampling oscilloscope (Keysight N1000A) to record the eye diagram. In order to further verify the performance of the TCO electrode modulator for high-speed data transmission, a range of tests were conducted. Firstly, we tested the eye diagram and bit error rate of the modulator with a modulation area length of 5 mm. The optical eye diagrams of OOK modulation at different modulation rates (56 Gbit/s, 80 Gbit/s, and 112 Gbit/s) on a 5 mm device are shown in Figs. 7(a)-(c). Additionally, we tested the PAM4 signal at a rate of 224 Gbit/s (112G Baud) as shown in Figure 7(d)-(e). In further measurements of the back-to-back (B2B) bit-error rates (BERs), we used an adjustable optical attenuator (Joinwit JW8507) and a 90GHz bandwidth optical detector (FINISAR XPDV4120R) along with a 70 GHz real-time oscilloscope (Keysight UXR0704A) to record the bit error rate test data. The BER calculation was performed offline using a Matlab-based offline algorithm. The BER test results were evaluated around the soft-decision forward error coding (SD-FEC) threshold ($2\times10^{-2}$).

The results indicate that the dynamic extinction ratios of the eye diagrams for both 56Gbit/s OOK and 224 Gbit/s PAM4 signals are (5.34, 4.15, 2.81, 2.32, 1.98) (dB) as shown in Figure 7(a)-(e). In Figure 7(f), we conducted bit error rate tests on the 112 Gbit/s OOK and PAM4 signals, as well as the 224 Gbit/s PAM4 signal. It was observed that under the PAM4 signal transmitted at a high rate of 224 Gbit/s, the bit error rate (BER) can be reduced to below the hard decision forward error coding (HD-FEC) threshold ($3.8\times10^{-3}$) when the received optical power is greater than -13dBm.

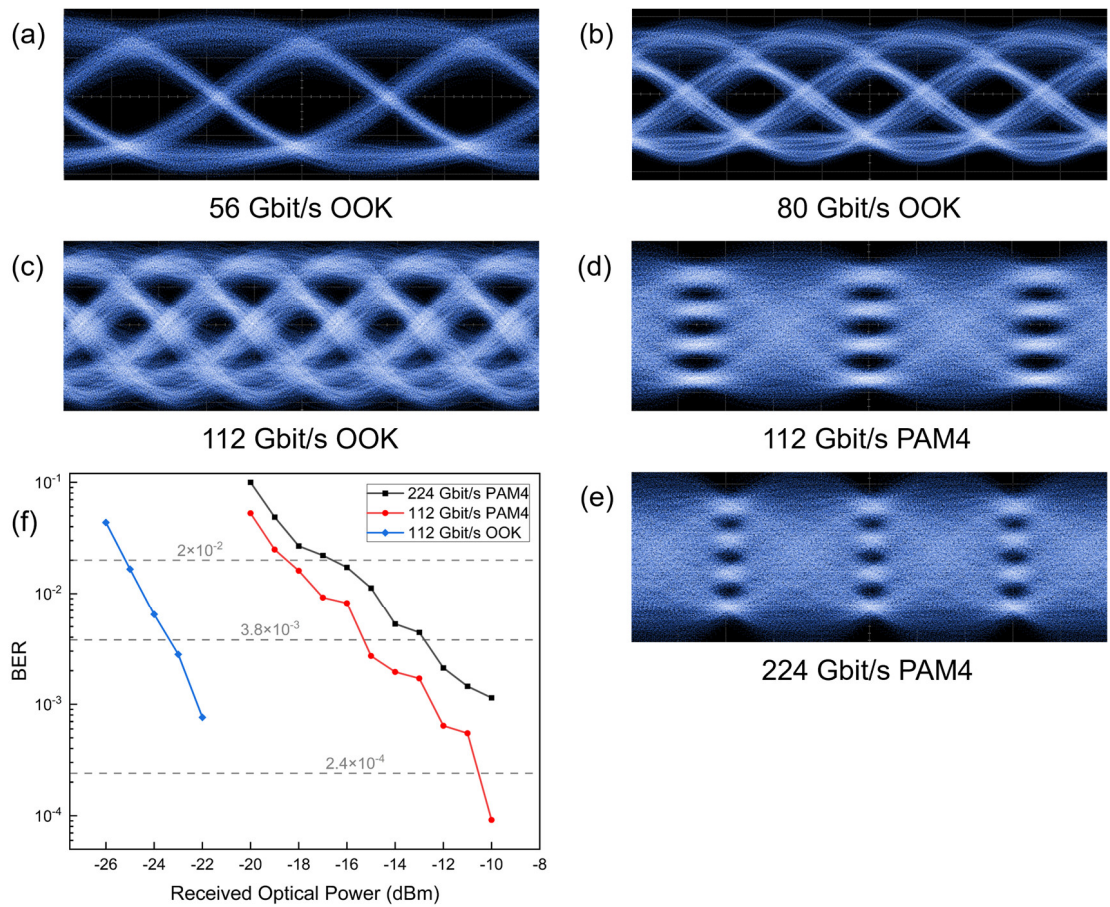

Fig. 7. High-speed eye diagram and bit error test results of 5mm devices. OOK format (a) 56 Gbit/s, (b) 80 Gbit/s, (c) 112 Gbit/s. PAM4 format (d) 112 Gbit/s (56 GBaud), (f) 224 Gbit/s (112 GBaud). (f) in different B2B bit error rate test results of 112 Gbit/s OOK, 112 Gbit/s PAM4, and 224 Gbit/s PAM4 measured under receiving power.

To further demonstrate the modulation performance enhancement achieved by introducing TCO

electrode materials, we compared the eye diagrams of two devices. The devices were two TFLN modulators with a modulation arm of 8 mm length, one with TCO loaded, and one without TCO loaded. The two devices were both operating in 56 Gbit/s OOK mode under the same driving voltage. The eye diagram results are shown below. The results clearly indicate a significant improvement in the modulator extinction ratio performance, increasing from 1.3dB to 6.0dB as a result of loading TCO.

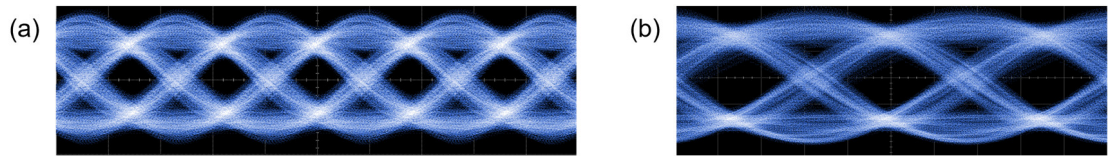

Fig. 8. 56 Gbit/s rate eye diagram test of 8mm device (a) without TCO loaded, (b) with TCO loaded.

## 4. Discussion

In this work, we introduced an ITO film into the electrode to effectively improve the modulation efficiency of the LNOI modulator, while maintaining its high modulation bandwidth advantage. However, during the measurement, we observed that the eye diagram extinction ratio of the modulator was not high, possibly due to asymmetric light absorption loss caused by a processing error in the metal electrode. Furthermore, the addition of TCO material to the electrode significantly improved the eye diagram extinction ratio by approximately 4.7dB, as evident from a comparison.

It is worth noting that methods of applying new materials or device structures to LNOI modulators, such as the introduction of $SiO_2$ buffer layers[16] and high-dielectric cladding materials, have also been reported[40]. To evaluate the performance of current devices, we compared them with other advanced devices in Table 1, focusing on modulation efficiency, modulation bandwidth, and insertion loss. As displayed in Table 1, our device achieved the highest modulation efficiency while still maintaining large modulation bandwidth with acceptable insertion loss.

Table 1. Comparison of device performance from different articles

| Year | Working Band | $V_\pi L$ (V·cm) | IL (dB) | BW (GHz) | Ref. |
|---|---|---|---|---|---|
| 2018 | C-Band | 2.2 | <0.5 | 100 | [19] |
| 2019 | C-Band | 2.2 | 2.5 | >70 | [18] |
| 2020 | O-Band | 2 | 16 | / | [41] |
| 2020 | C-Band | 2.47 | 1.8 | >67 | [21] |
| 2021 | C-Band | 1.7 | 17 | >67 | [17] |
| 2021 | C-Band | 1.75 | 0.7/cm | >40 | [22] |
| 2021 | C-Band | 2.3±0.2 | 1±0.5 | >50 | [24] |
| 2022 | C-Band | 2.35 | 6.5±0.5 | 110 | [25] |
| 2022 | C-Band | 2.2 | 0.2 | >67 | [26] |
| 2023 | C-Band | 1.41 | 0.5 | >40 | [40] |
| 2023 | O-Band | 1.85 | 2.5 | 65 | [42] |
| **2023** | **O-Band** | **1.024** | **2.9** | **108** | **This work** |

TFLN electro-optical modulator chips incorporating TCO electrodes demonstrate record-breaking high modulation efficiency, resulting in a 40% reduction in chip length compared to conventional gold electrode modulator chips for the same half-wave voltage requirement. This shorter phase-shift region not only leads to lower microwave losses but also contributes to a larger electro-optical modulation bandwidth. It is expected to lead to a nearly 70% increase in chip yield and a significant reduction in the cost of a single chip if our TCO electrode technology is employed in TFLN multi-channel optical transmitters in the future. This technology holds promising application prospects in data center optical modules and other relevant areas.

5. Conclusion

In this paper, a new type modulator based on LNOI platform is proposed. It utilizes combined electrodes made of TCO material and gold to achieve high electro-optic modulation efficiency. The spacing between the electrodes is only 3 μm, with 2.9 dB optical insertion loss, thanks to a 100nm buffer layer. As a result, the modulator exhibits an impressively low half-wave-voltage length product of 1.02 V·cm. The modulation region, which is 5 mm long, shows a EO 3dB transmission roll-off at 108 GHz. Successful demonstrations of on-off keying (OOK) modulation up to 112 Gbit/s and four-level pulse amplitude modulation (PAM-4) up to 224 Gbit/s are also shown. The proposed Thin-film Lithium Niobate Modulators Applied ITO Composite Electrode will significantly decrease the cost and power consumption of the TFLN based data center optical modules and short-distance coherent transmission modules.